\documentstyle[12pt]{article}
\begin{document}
\title{An Underpinning for Space Time}
\author{B.G. Sidharth\\Centre for Applicable Mathematics \& Computer
Sciences\\B.M. Birla Science Centre, Adarsh Nagar, Hyderabad - 500 063 (India)}
\date{}
\maketitle
\footnotetext{Email:birlasc@hd1.vsnl.net.in}
\begin{abstract}
We argue, that from a background pre space-time Zero Point Field, physical
space-time emerges on averaging over unphysical Compton scales.
\end{abstract}
\section{Introduction}
In a previous communication\cite{r1} it was shown how from the fractal
dimension two of a Quantum Mechanical path, the space time divide emerges: The
real coordinate $x$, shows up as a complex number $x + ix'$, and it was
deduced that $x' = ct$. The underpinning for the complex coordinate is the
double Weiner process, which in the Nelsonian formulation leads to a complex
velocity (Cf.ref.\cite{r1} and \cite{r2}). It also appears as zitterbewegung
and the resulting complex or non Hermitian position operator. It was also argued\cite{r3} that this Quantum Mechanical indeterminism gives birth to
time. Indeed it was pointed out that the above fractal behaviour has an immediate
analogue in Richardson's coastline\cite{r4}, while the Compton scale is the
thick brush that delineates the coastline\cite{r5}. We will now see that all this
implies the emergence of space-time from a pre space-time background Zero
Point Field or Quantum Vacuum.\\
\section{Space-time}
In the light of the above comments, we can now notice that within the Compton
time, for example we have a double Weiner process leading to non differentiability
with respect to time. That is, at this level time in our usual sense does
not exist. To put it another way, within the Compton scale we have the
zitterbewegung effects - these are non local and chaotic. As pointed out by
Dirac\cite{r6}, it is only after an averaging over these scales that meaningful
physics emerges. Indeed in the stochastic formulation of the Schrodinger,
Klein-Gordon and Dirac equations\cite{r5,r7}, we again encounter minimum space
time units, the Compton scale within which there are unphysical effects. It
is only outside the Compton scale that physics emerges.\\
This is a Quantum Mechanical and an experimental fact. It expresses the
Heisenberg Uncertainty Principle - space time points imply infinite momenta
and energies and are thus not meaningful physically. However Quantum Theory
has lived with this contradiction\cite{r8}. To put it simply to measure space
or time intervals we need units which can be to a certain extent and not
indefinitely subdivided - but already this is the origin of discreteness.
So physical time emerges at values greater than the minimum unit, which has
been shown to be at the Compton scale\cite{r9}. Going to the limit of space-time
points leads to the well known infinities of Quantum Field Theory (and classical
electron theory) which require renormalization for their removal.\\
The conceptual point here is
that time is in a sense synonymous with change, but this change has to be
tractable or physical. The non differentiability with respect to time, due
to the double Weiner process, within the Compton time, precisely highlights
time or change which is not tractable, that is is unphysical. However Physics,
tractability and differentiability emerge from this indeterminism once averages over the zitterbewegung
or Compton scale are taken. It is now possible to track time physically in
terms of multiples of the Compton scale.\\
To elaborate on this point it may be
mentioned that in\cite{r9,r10} and elsewhere, it was argued that given $n$
particles in the universe, within the Compton time $\tau, \sqrt{n}$ particles
would be fluctuationally or unphysically created (out of a background Zero
Point Field), so that we would have,
\begin{equation}
\frac{dn}{dt} = \frac{\sqrt{n}}{\tau}\label{e1}
\end{equation}
On integrating (\ref{e1}) from $n = 0$ to $n = N \sim 10^{80}$, as we have in
the present universe, it was shown that we recover the flow of time, infact
the correct age of the
universe, as indeed can be easily verified from (\ref{e1}).\\
The above considerations can be looked at from another point of view. Time is
essentially an ordering or sequencing of events. We would like to know the
basis on which this ordering takes place and which leads to our physical universe.
The question is, can we liberate this sequence of events from any ordering
at all, even though this would on the face of it lead to a totally chaotic
universe without any physics. However within the Compton scale this should
certainly be possible:\\
As noted in \cite{r9}, let us start with hypothetical instantaneous point particles
(or Zero Point Energies or Ganeshas as they were designated). Their states could
be denoted by $\phi_n$, which form a basis, so that a general state could be
written as
\begin{equation}
\psi = \sum_{n} c_n\phi_n,\label{e2}
\end{equation}
$\phi_n$ could be eigen states of for example the position or energy, with eigen values
$\lambda_n$. It is known that, owing to the Random Phase Axiom, viz.,
\begin{equation}
\overline{(c_n,c_m)} = 0, \quad n \ne m.\label{e3}
\end{equation}
where $n$ could stand not for a single state but also for a set of states
$n_i$ and so also $m$, and where the bar denotes an average over a suitable
interval we have\cite{r11}
\begin{equation}
\psi = \sum_{n} b_n\phi_n ,\label{e4}
\end{equation}
where $|b_n|^2 = 1$ if $\lambda < \lambda_n < \lambda + \Delta \lambda \quad \mbox{and} \quad = 0$
otherwise.\\
What all this means is the following. We take a totally random sequence of
states like $\phi_n$. Such a sequence averaged over the interval $\tau$, the
Compton time and there by using (\ref{e3}), constitutes a particle that
has physical existence, as expressed by (\ref{e4}). Without such an average,
that is within the Compton time, we have instead of equation (\ref{e4}), equation (\ref{e2}), wherein
there is no ordering or sequencing whatsoever, that is time in the physical
sense has no meaning. This is a pre time (or pre space time) perfectly random
background Zero Point Field scenario. It is only on averaging that we recover
the physically meaningful equation (\ref{e4}).\\
Identical arguments apply to the case of space, except that there is the space
time divide arising out of the fractal two dimensionality of the Quantum path as noted above:
We have $x + ict$. Indeed it was pointed out\cite{r1} that this is the origin
of Special Relativity. This apart we can proceed in terms of the Compton
wavelength $l$ to get this time instead of (\ref{e1}),
\begin{equation}
\frac{dn}{dx} = \frac{\sqrt{n}}{l}\label{e5}
\end{equation}
Integrating (\ref{e5}) exactly as we did (\ref{e1}), we recover this time, the
well known hitherto empirical Eddington formula,
$$R \sim \sqrt{N} l$$
(whose time analogue from (\ref{e1}) is, $T \sim \sqrt{N}\tau)$.\\
Once again within the Compton wavelength we have non local unphysical effects.\\
However as we will now see, in the case  of space, we recover three dimensions.
The reason is that as pointed out in\cite{r12,r13} and elsewhere the zitterbewegung
or unphysical effects represent charge and double connectivity or spin half which in terms of spin networks\cite{r14}
or similar arguments\cite{r15} immediately leads to three dimensionality. Thus the
the three dimensionality is not apriori, but rather is a holistic property arising
from several particles.\\
Another way of looking at this is when two particles interact without any specific
reference to electromagnetism and three dimensionality, their potential
energy is given by, as is well known\cite{r16}
\begin{equation}
V = -\alpha \frac{\hbar c}{r}\label{e6}
\end{equation}
where $\alpha$ is the fine structure constant and $r$ the distance between
the two particles, without reference to any dimensionality.\\
We would now like to point out that the significance of (\ref{e6}) is that
it gives the inverse square Coulumb Force Law and therefore three dimensionality.\\
Indeed as Barrow \cite{r8} points out "Interestingly the number of dimensions of space
which we experience in the large plays an important role.... it also ensures
that wave phenomena behave in a coherent fashion. Were there four dimensions
of space, then simple waves would not travel at one speed in free space.... in any
world but one having three large dimensions of space, waves would become
distorted as they travel..."
Also as pointed out in\cite{r17}, in the case of two elementary particles separated
by a distance $r$, as the spectral density of the background Zero Point Field
is given by
$$\rho (\omega )\propto \omega^3$$
where $\omega$ is $\propto \frac{1}{r}$, the electromagnetic force is
given by
$$\mbox{Force} \quad \propto \int^{\infty}_r \omega^3 d r, \propto \frac{1}{r^2}$$
which in effect is the same as (\ref{e6}).
\section{Discussion}
The picture that presents itself is the following: There is a perfectly random,
incoherent background Quantum Vacuum or Zero Point Field. From what we term as
fluctuations of this Quantum Vacuum, particles are created at the Compton
scale. This is a transition from an equation like (\ref{e2}) which
represents total incoherence, to (\ref{e4}), where we have particles occupying
a coherent space time with physics. The coherence or link between the various
particles is interaction provided by the background Zero Point Field within
the Compton scales, or in more conventional language by the virtual photons
linking the various constitutents. In the transition from (\ref{e2}) to (\ref{e4}),
there has been a totally random sequencing within the Compton scale, as
expressed by the Random Phase equation (\ref{e3}). In other words within the
Compton scale, there is no physics - indeed this is the Zitterbewegung region
of non local effects.


\begin{thebibliography}{99}
\bibitem {r1} B.G. Sidharth, "Space Time at a Random Heap", to appear in
Chaos Solitons and Fractals.
\bibitem {r2} L. Nottale, Chaos, Solitons \& Fractals, 4, 3, 361-388, 1994,
and references therein.
\bibitem {r3} Sidharth, B.G., "Comment on the Paper 'Unification of Fundamental....'",
to appear in Chaos Solitons and Fractals.
\bibitem {r4} B.B. Mandelbrot,"The Fractal Geometry of Nature", (1982) W.H. Freeman,
New York, pg.2,18,27.
\bibitem {r5} B.G. Sidharth, "The Chaotic Universe", CSF 11 (2000), pp.1171-1174.
\bibitem {r6} P.A.M. Dirac, {\it The Principles of Quantum Mechanics}, Clarendon
Press, Oxford, 1958, p263.
\bibitem {r7} B.G. Sidharth, "The Universe of Chaos and Quanta", CSF 11 (2000), pp1269-1278.
\bibitem {r8} J.D. Barrow, "Theories of Everything", Vintage, London, 1992.
\bibitem {r9} B.G. Sidharth, "Universe of Fluctuations", Int.J.of Mod.Phys. A
13(5), 1998, pp599ff.
\bibitem {r10} B.G. Sidharth, International Journal of Theoretical Physics,
37 (4), 1307-1312, 1998.
\bibitem {r11} K. Huang, "Statistical Mechanics", Wiley Eastern, New
Delhi, 1975.
\bibitem {r12} B.G. Sidharth, "Quantum Mechanical Black Holes:Towards a
Unification of Quantum Mechanics and General Relativity", IJPAP, 35, 1997.
\bibitem {r13} B.G. Sidharth, Gravitation \& Cosmology, Vol.4, No.2, 1998.
\bibitem {r14} R. Penrose, "Angular Momentum: An approach to combinational
space-time" in, "Quantum Theory and Beyond", Ed., Bastin, T., Cambridge
University press, Cambridge, 1971.
\bibitem {r15} C.W. Misner, K.S. Thorne and J.A. Wheeler, "Gravitation", W.H.
Freeman, San Francisco, 1973.
\bibitem {r16} T. Jacobsen, European Journal of Physics, \underline{17}, 1996, p.92.
\bibitem {r17} B.G. Sidharth, in Instantaneous Action at a Distance in
Modern Physics: "Pro and Contra" , Eds., A.E. Chubykalo et. al., Nova Science
Publishing, New York, 1999.
\end{thebibliography}
\end{document}